\documentclass[]{spie}  

\usepackage{amsmath,amsfonts,amssymb}
\usepackage{booktabs}
\usepackage{float}
\usepackage{graphicx}
\usepackage[colorlinks=true, allcolors=blue]{hyperref}
\usepackage{xcolor} 

\title{Tolerancing the PIAA-ZWFS: a practical and robust wavefront sensor that approaches the fundamental sensitivity limit}

\author[a]{Adam K. Taras}
\author[a]{Sebastiaan Y. Haffert}
\author[a]{Louis Desdoigts}
\affil[a]{Leiden Observatory, Leiden University, PO Box 9513, 2300 RA Leiden, The Netherlands}

\authorinfo{Further author information: (Send correspondence to A.K.T.)\\A.K.T.: E-mail: taras@strw.leidenuniv.nl}

\usepackage{acronym}		
\usepackage{glossaries}		

\newcommand{\fib}{fibre} 

\newacronym{USyd}{USyd}{the University of Sydney}
\newacronym{ANU}{ANU}{Australia National University}
\newacronym{ESO}{ESO}{European Southern Observatory}

\newacronym{NSF}{NSF}{National Science Foundation}
\newacronym{LIEF}{LIEF}{Linkage Infrastructure, Equipment and Facilities}

\newacronym{VLTI}{VLTI}{Very Large Telescope Interferometer}
\newacronym{ATs}{ATs}{auxiliary telescopes}
\newacronym{UTs}{UTs}{unit telescopes}
\newacronym{ELT}{ELT}{Extremely Large Telescope}
\newacronym{JWST}{JWST}{James Webb Space Telescope}
\newacronym{DCT}{LDT}{Lowell Discovery Telescope}
\newacronym{GMT}{GMT}{Giant Magellan Telescope}
\newacronym{LBT}{LBT}{Large Binocular Telescope}

\newacronym{STS}{STS}{six telescope simulator}
\newacronym{GPAO}{GPAO}{GRAVITY+ adaptive optics}

\newacronym{WFS}{WFS}{wavefront sensor}
\newacronym{ADC}{ADC}{atmospheric dispersion corrector}
\newacronym{LDC}{LDC}{longitudinal dispersion corrector}
\newacronym{AO}{AO}{adaptive optics}
\newacronym{SNR}{SNR}{signal to noise ratio}
\newacronym{ADU}{ADU}{arbitrary data units}
\newacronym{RMS}{RMS}{root mean square}
\newacronym{PIAA}{PIAA}{phase-induced amplitude apodization}
\newacronym{ZWFS}{ZWFS}{Zernike wavefront sensor}
\newacronym{GPU}{GPU}{graphics processing unit}
\newacronym{PAWS}{PAWS}{piston-adapted wavefront sensor}
\newacronym{MKID}{MKID}{microwave kinetic inductance detector}

\newacronym{OAP}{OAP}{off-axis paraboloid}
\newacronym{DM}{DM}{deformable mirror}
\newacronym{MEMS}{MEMS}{micro-electromechanical system}
\newacronym{IR}{IR}{infrared}
\newacronym{SLM}{SLM}{spatial light modulator}
\newacronym{OPD}{OPD}{optical path difference}

\newacronym{SMF}{SMF}{single-mode \fib{}}
\newacronym{MCF}{MCF}{multicore \fib{}}
\newacronym{MMF}{MMF}{multimode \fib{}}
\newacronym{SLD}{SLD}{superluminescent diode}

\newacronym{CAD}{CAD}{computer aided design}
\newacronym{GUI}{GUI}{graphical user interface}
\newacronym{BFGS}{BFGS}{Broyden–Fletcher–Goldfarb–Shanno}
 
\begin{document} 
\maketitle

\begin{abstract}
High-contrast imaging demands extremely sensitive wavefront sensing to correct atmospheric effects and surface errors. While the limits of the sensitivity of a wavefront sensor are well known, a practical, robust design that saturates these limits remains elusive. This work further investigates the PIAA-ZWFS (Phase-Induced Amplitude Apodization-Zernike Wavefront Sensor). In previous work, we developed a framework to optimise its design, maximising Fisher information per frame in the presence of phase aberrations. In these proceedings, we study the effect of various manufacturing and alignment errors in the system on the overall performance. We employ both traditional Monte Carlo sampling and a 2\textsuperscript{nd} order expansion using our auto-differentiable simulator. The performance of the PIAA-ZWFS is not significantly degraded by these errors at values typical for manufacturing.  
\end{abstract}

\keywords{Adaptive optics, wavefront sensing, differentiable design}

\section{INTRODUCTION}
Extreme adaptive optics~\cite{guyon2018extreme} is necessary for imaging rocky exoplanets around nearby star systems from the ground, where the aberrations induced by the atmosphere vary rapidly. Similarly, high-contrast direct imaging instruments from space will require picometer-level knowledge of the wavefront aberrations~\cite{steeves2020picometer}. In either case, we are particularly interested in gaining the most information about the state of the wavefront in a finite number of photons. The fundamental information limit for wavefront sensing has been known for some time~\cite{paterson2008towards}, however the most widely used architectures do not reach this limit. Even the \gls{ZWFS}\cite{zernike1934diffraction,n2013calibration}, widely regarded as the most sensitive, falls significantly short of this limit\cite{chambouleyron2023modeling}. 

Recent work\cite{taras2026differentiabledesignpiaazwfsflexible, haffert2023reaching} has proposed a more sensitive variant by including a set of \gls{PIAA} lenses with a modified \gls{ZWFS} mask, known as the PIAA-ZWFS. In these proceedings we begin a study of the tolerances of this wavefront sensor. Before conducting the Monte Carlo simulations that are typical for tolerancing optical systems, we also leverage the differentiability of the underlying code (implemented in $\partial$Lux\cite{desdoigts2023differentiable}) to compute the exact Hessian of the system, providing a more complete picture of which parameters matter the most.

\begin{figure}[h]
    \centering
    \includegraphics[width=0.99\linewidth]{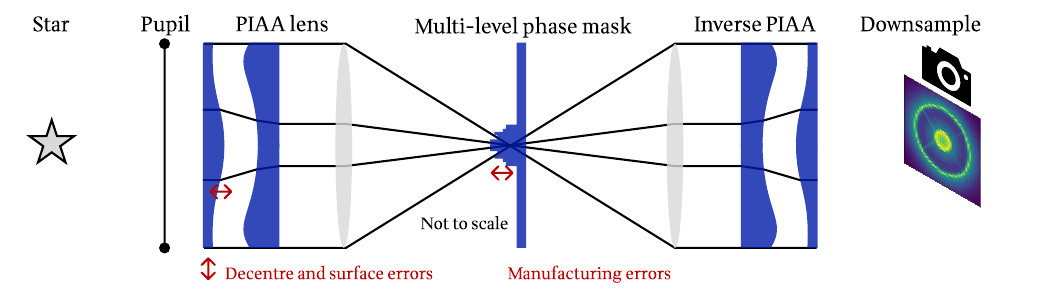}
    \caption{Schematic of the PIAA-ZWFS. After starlight passes through the pupil, a pair of PIAA lenses apodise the profile. When brought to focus, only a small mask is needed with a spatially varying glass thickness. After re-imaging, inverse PIAA lenses restore the geometry. In this work, we explore tolerances to errors in the system, such as surface form and decentre errors on the PIAA lenses, and height variations in the phase mask. }
    \label{fig:overview}
\end{figure}

\autoref{fig:overview} illustrates the PIAA-ZWFS and the degrees of freedom we explore here. We explore perturbations in the decentre of each PIAA lens (independent for all four), the form errors in the aspheres (assuming the forward and inverse lens in each pair is identical) and height errors in the phase mask (where each grayscale level is over or under etched). As we shall see, the inverse PIAA lenses do not matter much for the modes we used, and hence the assumption of identical manufacturing of lenses with the same static errors does not affect the final result. 

The Fisher information of such a system under a combination of photon and read noise is
\begin{align}\label{eq:fi}
    \text{FI}_{a_k a_l} = \sum_j \frac{(\partial_{a_k}I_j)(\partial_{a_l}I_j)}{I_j+\sigma_R^2},
\end{align}
where $a_k, a_l$ are the mode coefficients of phase aberrations in the pupil plane, $I_j$ is the intensity on the detector in pixel $j$, and $\sigma_R$ is the read noise of the detector. 

The loss function is 
\begin{align}
    \mathcal{L}(\theta) = \frac{I_{\mathrm {tot}}}{n_{\mathrm{modes}}} \mathrm{Tr}(\mathrm{FI}(\theta)^{-1}),
\end{align}
where $I_{\mathrm {tot}}$ is the total flux, $n_{\mathrm{modes}}$ is the number of modes sensed and $\theta$ is the design parameters of the wavefront sensor (\gls{PIAA} lens aspherical coefficients, phase mask heights). This is a typical loss function used in Bayesian experimental design and measures the average (per mode) variance of a maximum likelihood estimator in the high Strehl regime. We now turn to study how deviations of $\theta$ from the optimal value degrade $\mathcal{L}(\theta)$.

\section{TOLERANCES OF THE PIAA-ZWFS}

\subsection{TOLERANCES USING QUADRATIC APPROXIMATIONS}

In this work, we take a previously optimised design \cite{taras2026differentiabledesignpiaazwfsflexible} and evaluate the manufacturing and alignment tolerances required to achieve the performance. Specifically, we evaluate a PIAA-ZWFS for the MagAO-X pupil operating across the Z band (900\,nm centre wavelength, 152\,nm bandwidth). The aberration basis is the Fourier terms with only 9 terms total for computational reasons, with plans to extend it for the future. These frequencies still capture the behaviour of the idealised system for very high frequencies, but more work is needed to verify this is the case for systems with errors. 

At the optimal point, the Hessian of the loss function with respect to the parameters is the leading term in a Taylor expansion, as the gradients have vanished (by definition of an optimal point). Hence, the structure and relative size of values in the Hessian describes the sensitivity of the design to different errors. Values along the diagonal represent the curvature of the loss function in that axis in parameter space. Positive (negative) off-diagonal elements indicate that anti-correlated (correlated) errors will improve the loss function relative to if deviations only happened in one parameter. 

\begin{figure}[h]
    \centering
    \includegraphics[width=0.99\linewidth]{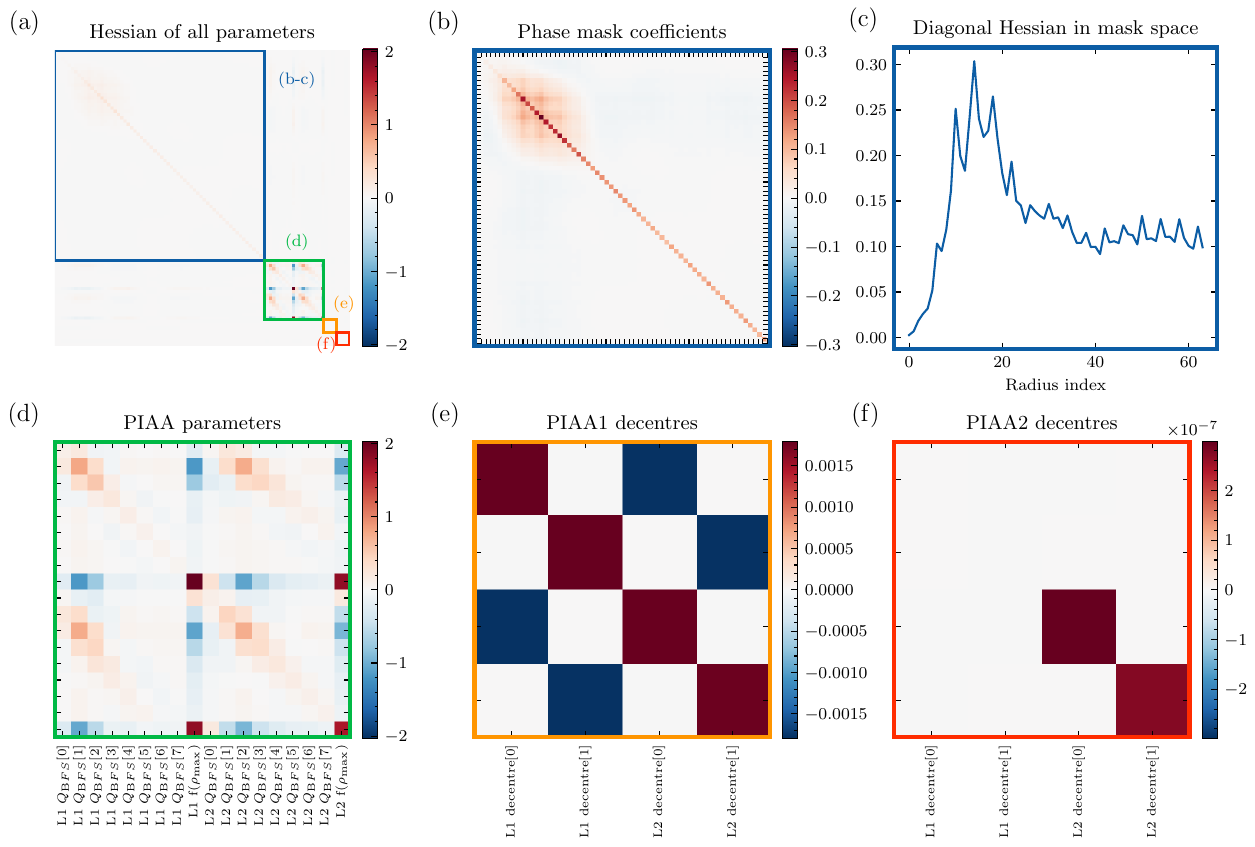}
    \caption{Hessian of the loss function evaluated at the local optimum, with different visualisations for each error in the system to highlight the different scales. (a) all parameters, (b) phase mask height errors in each annular ring, (d) PIAA lens form errors, (e-f) decentering errors on the input and inverse PIAA respectively. Values off the diagonal show cross-terms, with negative values indicating that positively correlated errors will cancel out to some extent. (c) visualises the diagonal of (b) as a function of radius.}
    \label{fig:hess}
\end{figure}

The Hessian of the loss function is visualised in \autoref{fig:hess}. It has units of loss ($\mu$m$^2$ per photon per frame/[parameter units]$^2$). In this case we keep all parameter units to be in $\mu$m, so in fact the Hessian simply has units of inverse photons. Whilst only accurate for small errors, this alone gives an idea of which parameters are the most significant and must be toleranced carefully. Subplot (a) illustrates the full Hessian. The largest term is the spherical term of the \gls{PIAA} lenses in (d), where deviations from the design result in a defocus on the phase mask. The positive off-diagonal elements indicate that a larger radius of curvature on the first lens can, in principle, be partially compensated by a lower radius of curvature on the second. Analysis of correcting manufacturing errors with alignment changes (such as varying the distance between the lenses) is left as future work. We also observe interesting trends in the tolerances of other components. For the phase mask, errors in levels for indices 10-20 seem to be more detrimental than elsewhere, and with larger off diagonal terms. Finally, the decentre tolerances are far more strict for the first PIAA lens pair than the inverse. This is expected as static phase aberrations from a flat wavefront lower the sensitivity of this sensor, but minor pupil re-mappings do not -- recall that the reason for the inverse PIAA is not sensitivity, but to recover the pupil geometry and avoid aliasing for high-frequency modes. In (e) we also observe off-diagonal terms nearly as large as the diagonal, with opposite sign. This indicates that decentres of one lens can be compensated by decentring the second lens in the same direction. 

In order to convert the Hessian into predictions of the distribution after manufacturing errors a probability distribution must be selected. If one assumes the errors on parameters are Gaussian, then it can be shown that the expectation value of the loss is approximately given by
\begin{align}\label{eq:2nd_order}
    \mathbb{E}_X[\mathcal{L}(X)] \approx \mathcal{L}(\theta_\mathrm{min}) + \frac{1}{2}\mathrm{Tr}\left(\left.H_\mathcal{L}\right|_{\theta_\mathrm{min}} \Sigma\right)
\end{align}
where $X\sim \mathcal{N}(\theta_\mathrm{min}, \Sigma)$ is the Gaussian random variable, and $H_\mathcal{L}$ is the Hessian of $\mathcal{L}$. This is true even away from the minimum, since the symmetry of the Gaussian distribution means that linear terms contribute nothing to the expectation value by symmetry. The above is a second order approximation to the size of the gap in the Jensen inequality -- that, for a convex function, the expectation of the function under a probability distribution is always greater than the function value at the expectation of the probability distribution. This 2\textsuperscript{nd} order expansion is also used in the following section as a comparison.

\subsection{TOLERANCES FROM MONTE CARLO METHODS}

We now turn to sampling a range of possible realisations of the PIAA-ZWFS. This Monte Carlo procedure involves all the parameters, however we restrict the size of the errors on all parameters to be scaled by a single value to understand how feasible a well made system would be. 

\autoref{tab:tols} gives the nominal values before any scaling. These represent typical ``precision'' quality specifications from asphere manufacturing and grayscale lithography, and a reasonable estimate for lens alignment when using motorised stages with an alignment sensor (either a target or a PSF camera). We anticipate that some values will be adjusted during the procurement stage but do not expect the conclusions to change. 

\begin{table}[h]
\centering
\caption{Nominal tolerances for the Monte Carlo simulations}
\label{tab:tols}
\begin{tabular}{@{}lcc@{}}
\toprule
Parameter           & \multicolumn{1}{l}{Nominal tolerance} & \multicolumn{1}{l}{Unit} \\ \midrule
PIAA sphere QBFS   & 50                                    & nm                       \\
PIAA spherical term & 0.1                                   & \%                       \\
PIAA lens decentre  & 1                                     & $\mu$m                       \\
Phase mask heights  & 30                                    & nm                       \\ \bottomrule
\end{tabular}
\end{table}

\begin{figure}[h]
    \centering
    \includegraphics[width=0.50\linewidth]{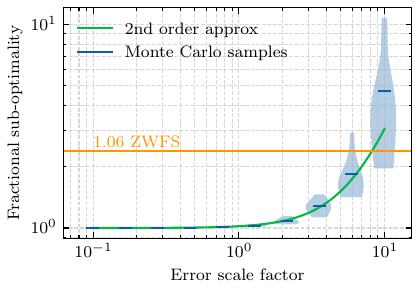}
    \caption{Monte Carlo simulation (blue) and 2\textsuperscript{nd} order expansion (green) estimate of the PIAA-ZWFS under different error scales, relative to the values given in \autoref{tab:tols}. The vertical axis is the loss value (average covariance of each phase mode) as a fraction of the error-free case. Horizontal line indicates the mean of the distribution, which closely matches the 2\textsuperscript{nd} order approximation, given by \autoref{eq:2nd_order}. Even errors a few times expectations would suffice to improve on the nominal \gls{ZWFS}.}
    \label{fig:mc_tol}
\end{figure}

\autoref{fig:mc_tol} illustrates the results of the Monte Carlo tolerance analysis as we scale the errors above. Each blue region shows the distribution of $\mathcal{L}(\theta)$ for 100 samples. We observe that the nominal tolerances are more than sufficient, and we could even tolerate a factor $\approx7\times$ worse errors before reaching the loss value of the \gls{ZWFS} with dot size $1.06\lambda/D$. Each distribution is long tailed (noting in particular the logarithmic y scale), which is expected as the resulting distribution of Gaussian inputs through a quadratic is a generalised chi-squared distribution. We also plot the predictions of the second order expansion in \autoref{eq:2nd_order}, which matches well for the region we are interested in. Once the Hessian is computed, all scales are computed extremely quickly --  a feature that would also enable exploration of many possible tolerances for each parameter individually.  

\section{CONCLUSIONS \& FUTURE WORK}
In these proceedings we have studied some of the manufacturing and alignment tolerances of the PIAA-ZWFS. We leveraged our auto-differentiable optical simulator to compute the exact Hessian of the system, finding that the most significant tolerances (when all parameters are in the same units) are the \gls{PIAA} lenses profiles. However, given typical manufacturing and alignment tolerances, we anticipate that the final performance will also depend on the decentres of the \gls{PIAA} lenses, as the typical scale ($\sim\mu$m) is much larger than the form errors expected in manufacturing ($\sim10$nm). Verification of the tolerance estimates for very high frequency modes is needed, particularly with the current extremely low impact of the inverse \gls{PIAA} alignment. 

For future work we envision several promising studies. We have shown that the 2\textsuperscript{nd} order approximations to $\mathcal{L}(\theta)$ remain valid for the tolerances we could reasonably achieve, and hence a manufacturing aware optimisation could be useful. In addition, a study of how the manufacturing errors above can be mitigated by modifying the optic positions would provide a more complete picture of the limits of this application. For example, form errors on the first \gls{PIAA} lens could perhaps be mitigated by changing the depth of the phase mask. Finally, a study of the tolerance requirements of other wavefront sensors would give a more complete picture when selecting one for a specific application. 

\acknowledgments 
 
The authors acknowledge funding from NWO VI.Vidi.233.144. This work was performed using the compute resources from the Academic Leiden Interdisciplinary Cluster Environment (ALICE) provided by Leiden University. Microsoft Copilot was used in developing the software for this work.

\bibliography{report} 
\bibliographystyle{spiebib} 

\end{document}